# Analytical Modelling of a Plucked Piezoelectric Bimorph for Energy Harvesting

Michele Pozzi

Manufacturing and Materials Department, School of Applied Sciences

Cranfield University, Bedfordshire, MK43 0AL, UK

e-mail: m.pozzi@cranfield.ac.uk

## *Abstract*

Energy harvesting (EH) is a multidisciplinary research area, involving physics, materials science and engineering, with the objective of providing renewable sources of sufficient power to operate targeted low-power applications. Piezoelectric transducers are often used for vibrational, inertial and direct movement EH. One problem is that, due to the stiffness of the most common material (PZT) and typically useful sizes, intrinsic resonant frequencies are normally high, whereas the available power is often concentrated at low frequencies. The aim of the plucking technique of frequency up-conversion, also known as "pizzicato" excitation, is to bridge this frequency gap. In this paper, the technique is modelled analytically. The analytical model is developed starting from the Euler-Bernoulli beam equations modified for piezoelectric coupling. A system of differential equations and associated initial conditions are derived which describe the free vibration of a piezoelectric bimorph in the last part of the plucking excitation, i.e. after release. The system permits the calculation of the mechanical response and the time evolution of the power generated and represents a ready-to-use modelling tool which will be specially useful for optimisation work.

## 1 Introduction

Energy harvesting (EH) is a hot topic at the moment, attracting interest from mechanical, electronic and system engineers, businesses, material scientists and physicists. With the purpose of scavenging energy from the environment for the most diverse range of applications (sensing, monitoring, automation, healthcare, …) a host of physical phenomena are investigated and exploited. Among the sources of environmental energy firstly harvested is the vibrational, with harvesting often done by piezoelectric transducers. Currently, the piezoelectric material of choice for macroscopic energy-harvesters is undoubtedly lead zirconate-titanate (PZT) in one of its many flavours (the properties of PZT can be tailored by adjusting its exact composition). The advantages of compactness and light-weight are arguably superior to many other energy harvesting technologies.

When used as EH material, the stiffness of piezoelectric ceramics is a source of problems in matching the mechanical impedance of the transducer with that of the rest of the system, including the environment that supplies the latent energy to be harvested. For this reason, the piezoelectric transducer normally favoured in EH is the PZT bimorph: similar to a bimetallic strip, it exploits the differential strain in two adjacent layers of PZT to produce displacement through bending when energized and *vice versa* produces electrical energy when bent. The resulting, usually cantilevered, beam is a much more compliant structure than a stack-type device.

Piezoelectric bimorphs which harvest energy from environmental vibrations still encounter a frequency mismatch problem: the former have high resonance frequencies, the latter have a power spectrum richer at lower frequencies. A traditional solution is the use of slenderer bimorphs and/or the attachment of a seismic mass at their end to reduce the system's fundamental frequency. A drawback is that the high fundamental frequency intrinsic to piezoelectric transducers is sacrificed, even though it is always desirable as, given that a small amount of energy can be converted in each cycle, higher frequencies mean the potential of higher power generation. The frequency mismatch issue is even more severe within the scope of human-based energy harvesting, given that human motions are intrinsically slow, never exceeding a few hertz (with the notable exception of vocalization). In wearable harvesters, the application of the traditional approach , i.e. longer bimorphs and additional mass, would lead to unacceptable encumbrance.

It is therefore clear the need for a technique capable of bridging between the high-frequency response of piezoelectric energy harvesters and the low-frequency input that is most often available in the environment and on the human body. The plucking, or "pizzicato", excitation technique addresses this issue.

## 2  Plucking: principle, applications and modelling

Plucking is made of three main phases. In the **approach phase**, the distance between a bimorph and a plectrum is reduced until they come into contact. Immediately follows the **loading phase**, during which both elastic elements are deflected, according to their mechanical compliance: mechanical energy is input in the system plectrum-bimorph in the form of strain energy. As the deflection progresses, the overlap between the two elements is reduced so that their contact area becomes gradually smaller, until contact is lost (**release point**). From this instant on, the bimorph **vibrates** at its resonance frequency as a cantilevered beam. During these vibrations, much of the elastic strain energy stored in the loading phase is converted by the direct piezoelectric effect into electrical energy and transferred to the external circuit; the remaining part is dissipated through various mechanisms, like air damping, dielectric losses and material internal damping. The outcome of plucking is frequency up-conversion, as by one single slow movement of the plectrum a large number of vibrations are produced at high frequency. The objective of the present paper is to formulate an analytical model of the release phase, in which most of the energy is produced (Pozzi and Zhu 2011). The benefits, compared to other modelling strategies such as Finite Elements, are the insight associated to analytical approaches and a reduced computational cost, specially valuable for optimisation work.

It is worthwhile to highlight the differences between plucking and another technique of energy harvesting: impact excitation  (Renaud et al. 2009; Umeda et al. 1996). From the physical point of view, impact involves the transfer of momentum; mathematically, the system's initial conditions have non-zero velocity; the principle is similar to hammers striking the wires in a piano. In plucking excitation, the piezoelectric devices are slowly deformed and then released; mathematically, the system's initial conditions feature a non- zero displacement; the principle is equivalent to the plucking of chords in a guitar (or a violin when the *pizzicato* technique is used).

Frequency up-conversion techniques can be used in situations where the vibrational energy available is at much lower frequency than the transducers' natural frequency. For example, Kulah and Najafi (Kulah and Najafi 2004; Kulah and Najafi 2008) have developed a micro-scale electromagnetic harvester where the low-frequency vibration of a suspended magnet excites the high frequency vibration of a number of coil-carrying cantilevers placed around it. Frequency up-conversion techniques can also be applied to scavenge energy from slowly rocking platforms or slow rotary motions (Rastegar and Murray 2010) or also from oceanic waves (Murray and Rastegar 2009). The plucking method can also be applied to wearable energy harvesters, as human movements are always very low-frequency (Pozzi and Zhu 2011). In the same paper, a Finite Elements model was presented and validated by comparison with experimental results.

Although the principle behind the plucking technique of frequency up-conversion is rather simple, it is important to carry out some modelling of the process to understand its potential and its dominating features as an energy harvesting technique. For example, it is important to know the maximum force needed to fully deflect the bimorph; failing this, depending on the mechanical configuration, the plucking action may not be completed (if the mechanism providing the input energy stalls at a limiting force) or the process not exploited to the full (if the plectra are too compliant and contact is lost before the bimorph is deflected sufficiently). In fact, this force can be readily calculated with good accuracy from the equations describing the static deflection of a cantilever.

Another piece of information of great importance in the design of a "pizzicato" energy harvester is the optimal plucking frequency. For this, it is necessary to develop a dynamic, time-dependent description of the cantilever's movement. One can then calculate the time evolution of the energy produced after a plucking action; this is used to decide when sufficient energy has been extracted from that event and it is more beneficial to produce a new plucking rather than wait for a full conversion of the stored energy. The dynamic model reported here offers the information to answer this type of questions.

# 3   *Analytical modelling*

In this section, the classical treatment of a vibrating clamped-free beam is adapted to a plucked piezoelectric bimorph; for more details on the beam-related derivations, the reader is referred to a book on vibrations such as (Ferrari and Gatti 2012)

The governing differential equation for a piezoelectric bimorph[1] (Figure 1) is:

$$\frac{\partial^2 M}{\partial x^2} + c_a \frac{\partial u}{\partial t} + c_s I \frac{\partial^5 u}{\partial x^4 \partial t} + m \frac{\partial^2 u}{\partial t^2} = f(x,t) \qquad 1$$

where $c_s$ is the coefficient of internal damping, $c_a$ is the coefficient of viscous damping (air), $m$ is the linear mass density, $M$ is made of two terms, $M_{mech}$ and $M_{el}$, which are the bending moments of mechanical and electrical origin, respectively:

$$M = M_{mech} + M_{el} = B \frac{\partial^2 u}{\partial x^2} + \theta V \qquad 2$$

where the $\theta$ will be derived later and the bending stiffness is:

$$B = 2 Y_{pz} I_{pz} + Y_{ms} I_{ms} = 2 Y_{pz} \left[ \frac{w h_{pz}^3}{12} + w h_{pz} \left( \frac{h_{pz} + h_{ms}}{2} \right)^2 \right] + Y_{ms} \frac{w h_{ms}^3}{12} \qquad 3$$

here the *pz* and *ms* subscripts stand for piezoelectric layer and metal shim, respectively; $Y$ is the Young's modulus in the *x* or *y* direction[2]; $I$ is the second moment of an area (the parallel axis theorem was applied to calculate the $I$ for the two piezoelectric layers); $w$ is the width of the bimorph.

---

1   It is assumed that the geometrical parameters of the bimorph permit the use the Euler-Bernoulli beam approximation.
2   PZT ceramics are transversely isotropic as the poling direction is different from the other two, which are equivalent (the symmetry of a standing cone).

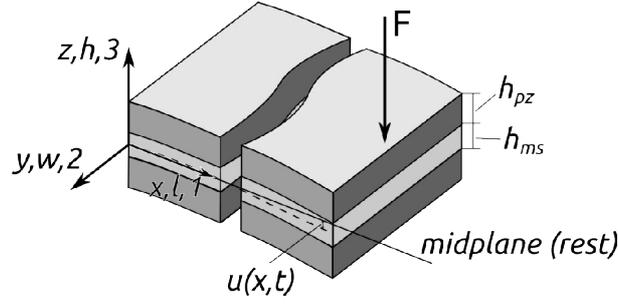

*Figure 1. Sketch of the piezoelectric bimorph with relevant dimensions and notations used in the analytical model*

The parameter $\theta$ couples the mechanical and the electrical fields and can be derived from the following constitutive equations of piezoelectricity, which gives the stress[3] $T$ as a function of the applied electric field:

$$T_1 = c_E S_1 - d_{31} Y_{pz} E .$$  4

We ignore its first term as it is already taken care of by equation 3; the second term originates:

$$M_{el} = 2 \int z\, da\, T = 2 \int_{h_{ms}/2}^{h_{ms}/2 + h_{pz}} z\, (dz\, w)\, d_{31} Y_{pz} \frac{V}{2 h_{pz}} = \frac{w\, d_{31} Y_{pz} (h_{pz} + h_{ms})}{2} V = \theta V$$  5

where we have used the relationship between electric field E and voltage V ( $E = -V/2 h_{pz}$ ), which assumes a series bimorph.

Once a plucked bimorph is released, it is not subjected to any external force, so we can set *f(x,t)=0*. Substitution of equation 2 into 1 followed by the normalisation of the longitudinal coordinate (we define $\xi = x/L$ ), gives:

$$\frac{B}{m L^4} \frac{\partial^4 u}{\partial \xi^4} + \frac{c_a}{m} \frac{\partial u}{\partial t} + \frac{c_s I}{m L^4} \frac{\partial^5 u}{\partial x^4 \partial t} + \frac{\theta}{m L^2} \frac{\partial^2 V}{\partial \xi^2} + \frac{\partial^2 u}{\partial t^2} = 0$$  6

Without introducing an explicit dependence of *V* on *x*, this term would disappear in the following manipulations; so we use the Heaviside step function (Erturk and Inman 2008) to state the fact that *V* is constant over the electrodes, which we assume extend over the whole bimorph, and zero outside:

$$V(\xi, t) = V(t)[H(\xi) - H(\xi - 1)]$$  7

We can now begin solving equation 6 by separation of variables, assuming that the solution is given by:

$$u(\xi, t) = \sum_{n=1}^{\infty} \phi_n(\xi) q_n(t)$$  8

where the spatial components $\varphi_n$ are an orthonormal base in the functional space $L^2$ ([0,1]) and can be chosen, as for a standard Euler-Bernoulli beam, as

$$\phi_n(\xi) = \cosh(k_n \xi) - \cos(k_n \xi) - \sigma_n [\sinh(k_n \xi) - \sin(k_n \xi)]$$  9

where $k_n$ are the solutions to 1 + cos k cosh k = 0 and

$$\sigma_n = \frac{\sinh(k_n) - \sin(k_n)}{\cosh(k_n) + \cos(k_n)}$$  10

Now equations 7, 8 and 9 are substituted into equation 6, then we calculate the internal product with a single $\varphi_m$ for every $m$ (in other words, we project onto every vector in the functional vector space); in the process we use the orthonormal relationship:

$$\int \phi_n \phi_m = \delta_{mn}$$

and the derived:

$$\int \phi_n \frac{\partial^4 \phi_m}{\partial \xi^4} = k_m^4 \delta_{mn}$$

---

3  We are here adhering to the convention, common in piezoelectricity, of naming T the stress and S the strain.

After the projection, we obtain (with the compact dot-notation for time derivatives):

$$\ddot{q}_n + 2\zeta_n \omega_n \dot{q}_n + \omega_n^2 q_n - \chi_n V = 0 \qquad 11$$

with

$$\omega_n = \frac{k_n^2}{L^2} \sqrt{\frac{B}{m}} \qquad 12$$

$$\zeta_n = \frac{c_a}{2m\omega_n} + c_s \frac{I}{2mL^4 \omega_n} \quad k_n^4 = \frac{c_a}{2m\omega_n} + c_s \frac{I}{2B} \omega_n$$

$$\chi_n = \frac{d_{31} Y_{pz} w (h_{pz} + h_{ms})}{2mL^2} \frac{\partial \phi_i(\xi=1)}{\partial \xi}$$

Equation 11 is essentially the same that would be found for a standard Euler-Bernoulli beam, with the addition of the last term, which represents the converse piezoelectric coupling, as it states that the voltage on the electrodes acts as a force to determine the time evolution of the mode shape.

The direct piezoelectric effect is still to be included. To take into account the effect of the electrical circuit connected to the bimorph, we have to relate the voltage to the current, which is the time derivative of the charge $Q$, which in turn is given by the surface integral of the electric displacement $D$ over the electrode area:

$$I(t) = \dot{Q}(t) = \frac{d}{dt} \int_0^1 D_3(\xi,t) w L d\xi = -\frac{d_{31} Y_{pz}(h_{ms}+h_{pz})w}{2L} \int_0^1 \frac{\partial^3 u}{\partial t \partial \xi^2} d\xi - \frac{wL\varepsilon_{33}^S}{2h_{pz}} \dot{V}(t) \qquad 13$$

Here $D_3$ is given by another constitutive equation of piezoelectricity:

$$D_3(x,t) = d_{31} Y_{pz} S_1(x,t) - \varepsilon_3^S \frac{V(t)}{2h_{pz}} \qquad 14$$

where we have focussed on $D_3$, which is parallel to the polarisation and so gives the charge density on the electrodes.

After the substitution, we decompose $u(x,t)$ as in expression 8, so that we find (using the dot-notation for time derivatives):

$$I(t) = -\frac{d_{31} Y_{pz} w (h_{ms}+h_{pz})}{2L} \sum_n \dot{q}_n \frac{d\phi(\xi=1)}{d\xi} - \frac{\varepsilon_3^S w L}{2h_{pz}} \dot{V}(t)$$

Combining this with Ohm's law for a resistor $R$ placed across the electrodes of the bimorph, we have:

$$\dot{V} + \frac{V}{\tau} + \sum_{n=1} \Phi_n \dot{q}_n = 0$$

with

$$\Phi_n = \frac{d_{31} Y_{pz} h_{pz} (h_{pz} + h_{ms})}{\varepsilon_3^S L^2} \frac{\partial \phi_n(\xi=1)}{\partial \xi}$$

$$\tau = \frac{\varepsilon_3^S R w L}{2 h_{pz}}$$

Substituting $s = \dot{q}$ to reduce the order, the system of differential equations to be solved can be written, for an arbitrary number $n$ of modes:

$$\begin{aligned}
\dot{q}_1 &= s_1 \\
\dot{q}_2 &= s_2 \\
&\ldots \\
\dot{q}_n &= s_n \\
\dot{s}_1 + 2\zeta_1 s_1 + \omega_1^2 q_1 - \chi_1 V &= 0 \\
\dot{s}_2 + 2\zeta_2 s_2 + \omega_2^2 q_2 - \chi_2 V &= 0 \\
&\ldots \\
\dot{s}_n + 2\zeta_n s_n + \omega_n^2 q_n - \chi_n V &= 0 \\
\dot{V} + \frac{V}{\tau} + \sum_{i=1}^{n} \Phi_i \dot{q}_i &= 0
\end{aligned} \qquad 15$$

This is to be solved with initial conditions that take into account the fact that at the time of release the bimorph is deflected. For simplicity, we assume that the bimorph starts from a static condition; therefore: its shape is that of a statically deflected cantilever, all time derivatives are zero and the voltage across its electrodes is zero, as it would have fully discharged across the resistor since it was deflected. Mathematically:

$$\begin{aligned}
q_i(0) &= \int_0^1 \phi_i(\xi) \frac{F L^3}{2 B} \xi^2 \left(1 - \frac{\xi}{3}\right) d\xi \quad \forall \quad i = 1..n \\
s_i(0) &= 0 \quad \forall \quad i = 1..n \\
V(0) &= 0
\end{aligned} \qquad 16$$

As evident from the formulation of the initial conditions, the force on the tip of the bimorph is inserted, using formulae for the static deflection of cantilevers, to give the initial deflection.

Equations 15 and 16, together with the symbols definitions given earlier can be directly used in mathematical software applications such as Maple® or Mathematica®, to solve the system for the desired number of mode shapes.

## 4    *General considerations and limitations*

The model includes only viscous damping within the material as a form of energy dissipation. As a result, conversion efficiency may be too high. A real EH faces other forms of energy loss, such as internal material dissipation within the plectra and friction in the bearings and in other mechanisms which are needed to transfer external energy into the harvester. These losses are device-specific and should be dealt with as part of the detailed design of each energy harvester.

Although the total energy generated increases with the time interval in which the bimorph is free to vibrate, the average power clearly decays hyperbolically[4] over long periods, giving rise to an optimal plucking frequency. Normally, the designer has some scope to determine the plucking frequency during operation, which should be optimised in the light of such considerations.

The model developed here only describes the release phase. The energy produced during loading is negligible at low deflection speeds, but becomes more important at high speeds (Pozzi and Zhu 2011).

---

4   The average power is given by $P_{av} = E(t)/t$, where $E(t)$ tends to an asymptotic value for $t \to \infty$.

Complex circuitry for electrical power management, such as SSHI in one of its variants (Guyomar and Lallart 2011), are very difficult to model analytically and have not been included (the assumption is that the bimorph is connected to a simple resistor). Their non-linear behaviour would alter the mechanical response of the bimorph to some extent, due to the coupling between electrical and mechanical fields introduced by piezoelectricity.

# 5 Conclusions and outlook

There is currently great focus within the EH community on broadband harvesters, which are not constrained to operate in a very narrow frequency range. Frequency up-conversion by plucking is offered as a solution to the frequency mismatch problem: albeit not suitable for all applications and environments, it offers promising results in different contexts.

The paper has presented a framework for analytical modelling of plucked bimorphs that can be used as is or adapted for other specific configurations. Beyond sizing of harvester and bimorphs, choice of materials and spacing of plectra, the designer must pay great attention to the interaction between plectra and bimorphs: choose the right materials, shape of plectra to attain the cleanest release (Pozzi&Zhu 2012) and minimisation of the wear of the parts in contact. Following this line of thoughts, other interaction mechanisms may deserve investigation, e.g. where the plectra are replaced with permanent magnets.

In conclusion, energy harvesting by plucking is still in its infancy and significant improvements in the technique and in the performance of the harvesters can be expected in the near future. This work aims at facilitating the development of improved harvesters by offering a ready-to-use analytical modelling framework that can be incorporated in optimisation routines.

## Acknowledgements

This work was made possible by the financial support of dstl, part of the Ministry of Defence of the UK, via the Engineering and Physical Sciences Research Council (EPSRC), grant no. EP/H020764/1. I also acknowledge Dr. Meiling Zhu for her support of this work as leader of the Energy Harvesting Research Group in Cranfield.